\documentclass{article}

\usepackage{preprint}
\usepackage[dvips]{epsfig}

\def\kbar{\overline{K}{}^{\,0}}
\def\dbar{\overline{D}{}^{\,0}}
\def\bbar{\overline{B}{}^{\,0}}
\def\cp{$CP$}
\def\cpv{$CPV$}
\def\ra{\!\rightarrow\!}

\def\simge{\mathrel{%
   \rlap{\raise 0.511ex \hbox{$>$}}{\lower 0.511ex \hbox{$\sim$}}}}
\def\simle{\mathrel{
   \rlap{\raise 0.511ex \hbox{$<$}}{\lower 0.511ex \hbox{$\sim$}}}}

\begin{document}

\title{\boldmath 
\vskip-0.75in
\flushright{\rm UCHEP-06-02}
\vskip0.50in
\centering{SEARCHES FOR $D^0$-$\dbar$ MIXING: \\
FINDING THE (SMALL) CRACK IN THE STANDARD MODEL}
}
\author{
A.\,J.\,Schwartz \\
{\em University of Cincinnati, P.O. Box 120011, Cincinnati, Ohio 45221} 
}
\maketitle
\baselineskip=11.6pt
\begin{abstract}
We review results from searches for mixing and \cp\ 
violation in the $D^0$-$\dbar$ system. No evidence 
for mixing or \cp\ violation is found, and limits 
are set for the mixing parameters $x$, $y$, $x'$, $y'$, 
and several $CP$-violating parameters.
\end{abstract}

\baselineskip=14pt

\section{Introduction}

Despite numerous searches, mixing between $D^0$ and 
$\dbar$ flavor eigenstates has not yet been observed. 
Within the Standard Model (SM), the short-distance ``box'' 
diagram (which plays a large role in $K^0$-$\kbar$ and 
$B^0$-$\bbar$ mixing) is doubly-Cabibbo-suppressed (DCS)
and GIM-suppressed; since the
$D^0$ decay width is dominated by Cabibbo-favored (CF)
amplitudes, $D^0$-$\dbar$ mixing is expected to be a rare 
phenomenon. Observing mixing at a rate significantly 
above the SM expectation could indicate new physics.

The formalism describing $D^0$-$\dbar$ mixing 
is given in several papers.\cite{xing,cicerone} 
The parameters used to characterize mixing are
$x=\Delta m/\overline{\Gamma}$ and
$y=\Delta\Gamma/(2\overline{\Gamma})$, where
$\Delta m$ and $\Delta \Gamma$ are the mass and decay 
width differences between the two mass eigenstates, 
and $\overline{\Gamma}$ is the mean decay width. Within 
the SM, $x$ and $y$ are difficult to calculate as there 
are long-distance contributions. 
For $m^{}_q\!\gg\!\Lambda^{}_{\rm QCD}$, these contributions 
can be estimated using the heavy-quark expansion;
however, $m^{}_c$ may not be large enough for this
calculation to be reliable.
Current theoretical predictions\cite{petrov} span a wide 
range: $|x|\!\sim\!|y|\!\sim$\,($10^{-7}$ to $10^{-2}$),
with the majority being $\simle 10^{-3}$.

For decay times $t\ll 1/\Delta m,\,1/\Delta\Gamma$,
which is well-satisfied for charm decay, the time-dependent 
$D^0(t)\ra f$ and $\dbar(t)\ra\bar{f}$ decay rates are
\begin{eqnarray}
\!\!R^{}_{D^0}\!\! & \!\!\!=\!\!\! & 
|{\cal A}^{}_f|^2\,
e^{-\overline{\Gamma}t} \left[\,1 + 
\left[\,y\,{\rm Re}(\lambda) - x\,{\rm Im}(\lambda)\,\right]
(\overline{\Gamma}t) + |\lambda |^2
\frac{(x^2 + y^2)}{4}(\overline{\Gamma}\,t)^2\,\right] 
\label{eqn:master1} \\
 & & \nonumber \\
\!\!R^{}_{\dbar}\!\! & \!\!=\!\!\! & 
|\bar{\cal A}^{}_{\bar{f}}|^2\,
e^{-\overline{\Gamma}t} \left[\, 1 +  
\left[\,y\,{\rm Re}(\bar{\lambda}) - 
x\,{\rm Im}(\bar{\lambda})\,\right]
(\overline{\Gamma}t) + |\bar{\lambda} |^2
\frac{(x^2 + y^2)}{4}(\overline{\Gamma}\,t)^2\,\right]\!,
\label{eqn:master2} 
\end{eqnarray}
where $\lambda=(q/p)(\bar{\cal{A}}^{}_f/{\cal A}^{}_f)$,
$\bar{\lambda}=(p/q)({\cal A}^{}_{\bar{f}}/\bar{{\cal A}}^{}_{\bar{f}})$,
$q$ and $p$ are complex coefficients relating flavor
eigenstates to mass eigenstates, and 
${\cal A}^{}_f\,(\bar{{\cal A}}^{}_f)$ and
${\cal A}^{}_{\bar{f}}\,(\bar{{\cal A}}^{}_{\bar{f}})$ 
are amplitudes for a pure $D^0$ ($\dbar$) state to
decay to $f$ and $\bar{f}$, respectively. 

In this paper we discuss five methods used to search for 
$D^0$-$\dbar$ mixing and \cp\ violation (\cpv). These 
methods use the following decay modes:\footnote{Charge-conjugate 
modes are implicitly included throughout this paper unless 
noted otherwise.}
semileptonic $D^0\ra K^+\ell^-\nu$ decays,
decays to \cp-eigenstates $K^+K^-$ and $\pi^+\pi^-$,
DCS $D^0\ra K^+\pi^-$ decays, 
$D^0\ra K^0_S\,\pi^+\pi^-$ decays, 
and multi-body DCS $D^0\ra K^+ n(\pi)$ decays. A newer
method based on quantum correlations\cite{asner_quantumcorr} 
in $e^+e^-\ra\psi''(3770)\ra D^0\dbar$ production is not 
discussed here. The flavor of a $D^0$ when produced 
is determined by requiring that it originate from a 
$D^{*+}\ra D^0\pi^+_s$ decay;
the charge of the low momentum (``slow'') $\pi^+_s$
determines the charm flavor at $t\!=\!0$.
As the kinetic energy released in $D^{*\,+}\ra D^0\pi^+_s$ 
decays is only 5.8~MeV (very near threshold), requiring that 
$Q\equiv M^{}_{K\pi\pi^{}_s}\!-M^{}_{K\pi}\!-m^{}_\pi$ be 
small greatly reduces backgrounds.

\section{\boldmath $D^0(t)\ra K^{(*)+}\ell^-\nu$ Semileptonic Decays}

Because the $K^{(*)\,+}\ell^-\nu$ final state can only be 
reached from a $\dbar$ decay, observing 
$D^0(t)\ra K^{(*)\,+}\ell^-\nu$ would provide clear
evidence for mixing. In Eq.~(\ref{eqn:master1}) only the 
third term is nonzero; integrating this term over all times 
and assuming $|q/p|=1$ (i.e., neglecting \cpv\ in mixing) gives
\begin{eqnarray}
\frac{\int R(D^0\rightarrow K^+\ell\nu)\,dt}
{\int R(D^0\rightarrow K^-\ell\nu)\,dt} 
& \!\!\approx\!\! & \frac{x^2+y^2}{2}\ \equiv\ r^{}_D\,.
\end{eqnarray}

Several experiments\cite{semi_all,semi_belle} have used this method 
to constrain $r^{}_D$\,; the most stringent constraint is from the 
Belle experiment using 253~fb$^{-1}$ of data.\cite{semi_belle}
Due to the neutrino, the final state is not fully reconstructed;
however, at an $e^+e^-$ collider there are enough kinematic 
constraints to infer the neutrino momentum. Specifically, 
momentum conservation prescribes 
$P^{}_\nu=P^{}_{CM}-P^{}_{\pi^{}_s K\ell}-P^{}_{\rm rest}$,
where $P^{}_{CM}$ is the four-momentum of the $e^+e^-$ center-of-mass
(CM) system, $\pi^{}_s,\,K$, and $\ell$ are daughters from
$D^*\ra D^0\pi^{}_s\ra\pi^{}_s K\ell\nu$, and $P^{}_{\rm rest}$ is
the four-momentum of the remaining particles in the event.
In the Belle analysis the magnitude $|P^{}_{\rm rest}|$ is
rescaled to satisfy $(P^{}_{CM}-P^{}_{\rm rest})^2=m^2_{D^*}$,
and after this rescaling the direction of $\vec{\bf p}^{}_{\rm rest}$ 
is adjusted to satisfy $P^2_\nu\,(=m^2_\nu)=0$.

\begin{figure}[th]
\begin{center}
\epsfig{file=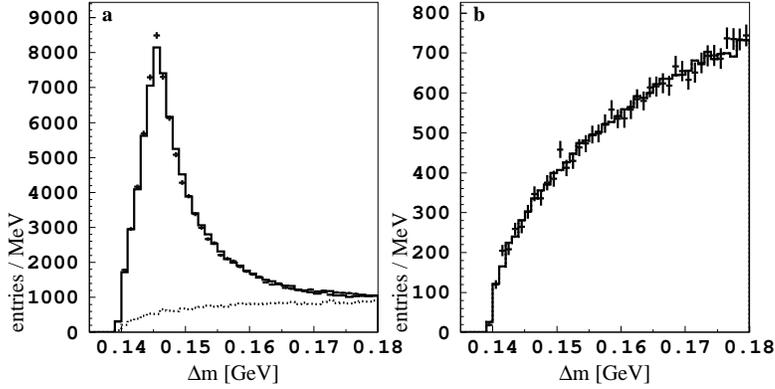,width=4.0in}
\end{center}
\vskip-0.25in
\caption{\it $\Delta M$ distributions for RS $D^0\ra K^-\ell^+\nu$ 
candidate decays (left) and WS $D^0\ra K^+\ell^-\nu$ candidate decays 
(right), from Belle using 253~fb$^{-1}$ of data.\cite{semi_belle} The 
WS plot shows no visible signal above background.
\label{fig:semi_belle}}
\end{figure}

The $\Delta M\!\equiv\!M^{}_{\pi^{}_s K\ell\nu}\!-M^{}_{K\ell\nu}$ 
distributions for ``right-sign'' (RS) $D^0\ra K^-\ell^+\nu$ 
and ``wrong-sign'' (WS) $D^0\ra K^+\ell^-\nu$ samples
are shown in Fig.~\ref{fig:semi_belle}. 
Sensitivity to mixing is improved by utilizing information on the
decay time, which is calculated by projecting the $D^0$ flight 
distance onto the (vertical) $y$ axis:
$t=(M^{}_{D^0}/c)\times (y^{}_{\rm vtx}-y^{}_{\rm IP})/p^{}_y\,$.
This projection has superior decay time resolution, as the beam 
profile is only a few microns in $y$ and thus the interaction 
point ($y^{}_{\rm IP}$) is well-determined. Events satisfying 
$t\!>\!\tau^{}_{D^0}$ are divided into six $t$ intervals, and the 
event yields $N^{(t)}_{\rm RS}$ and $N^{(t)}_{\rm WS}$, acceptance ratio 
$\varepsilon^{(t)}_{\rm WS}/\varepsilon^{(t)}_{\rm RS}$, and resulting 
mixing parameter $r^{(t)}_D$ are calculated separately for each.
$N^{(t)}_{\rm RS}$ and $N^{(t)}_{\rm WS}$ are obtained from fitting 
the $\Delta M$ distributions. Doing a $\chi^2$ fit to the six 
$r^{(t)}_D$ values gives an overall result
$r^{}_D=[0.20\pm 0.47\,({\rm stat})\pm 0.14\,({\rm syst})]\times 10^{-3}$,
or $r^{}_D\!\leq 0.10\%$ at 90\% C.L. 
No evidence for mixing is observed. The total number of signal
candidates in all $t$ intervals is $90601\pm 372$~RS events 
and $10\pm 80$~WS events.

\section{\boldmath $D^0(t)\ra K^+K^-,\,\pi^+\pi^-$ \cp-Eigenstate Decays}

When the final state is self-conjugate, e.g., $K^+K^-$, there 
is no strong phase difference between $\bar{\cal A}^{}_f$ and 
${\cal A}^{}_f$. Assuming $|\bar{\cal A}^{}_f|=|{\cal A}^{}_f|$
(no direct \cpv), $\lambda=-|q/p|\,e^{i\phi}$ and 
$\bar{\lambda}=-|p/q|\,e^{-i\phi}$, where $\phi$ is
a weak phase difference and the leading minus sign 
is due to the phase convention 
{\boldmath $CP$}$|D^0\rangle\!=\!-|\dbar\rangle$. Inserting 
these terms into Eqs.~(\ref{eqn:master1}) and (\ref{eqn:master2}) 
and dropping the very small last term gives
\begin{eqnarray}
R(D^0\ra K^+K^-) & = & 
|{\cal A}^{}_{K^+K^-}|^2\,e^{-\overline{\Gamma}t}
\left[\,1-\left|\frac{q}{p}\right|(y\cos\phi-
x\sin\phi)\overline{\Gamma}t\,\right] \nonumber \\
 & & \nonumber \\
 & \approx & |{\cal A}^{}_{K^+K^-}|^2\,e^{-\overline{\Gamma}t}
\,e^{-|q/p|(y\cos\phi-x\sin\phi)\overline{\Gamma}t} 
\label{eqn:kkmaster1} \\
R(\dbar\ra K^+K^-)  & \approx & 
|{\cal A}^{}_{K^+K^-}|^2\,e^{-\overline{\Gamma}t}
\,e^{-|p/q|(y\cos\phi+x\sin\phi)\overline{\Gamma}t}\,.
\label{eqn:kkmaster2}
\end{eqnarray}
Eqs.~(\ref{eqn:kkmaster1}) and (\ref{eqn:kkmaster2})
imply that the measured $D^0$ and $\dbar$ inverse lifetimes are 
$\overline{\Gamma}[1+|q/p|(y\cos\phi -x\sin\phi)]$ and
$\overline{\Gamma}[1+|p/q|(y\cos\phi +x\sin\phi)]$, respectively.
We define $y^{}_{CP}\equiv\tau^{}_{K^-\pi^+}/\tau^{}_{K^+ K^-}\!-1$,
which equals $|q/p|(y\cos\phi -x\sin\phi)$ for $D^0$ decays and
$|p/q|(y\cos\phi +x\sin\phi)$ for $\dbar$ decays.
For $|q/p|\!=\!1$, i.e., no \cpv\ in mixing, 
$y^{}_{CP}=y\cos\phi$ for equal numbers of $D^0$ 
and $\dbar$ decays together. If also
$\phi\!=\!0$ (no \cpv), $y^{}_{CP}=y$. The observable 
$y^{}_{CP}$ is measured by fitting the $D^0\ra K^+K^-$ 
and $D^0\ra K^-\pi^+$ decay time distributions.

\begin{figure}[th]
\begin{center}
\vskip0.10in
\epsfig{file=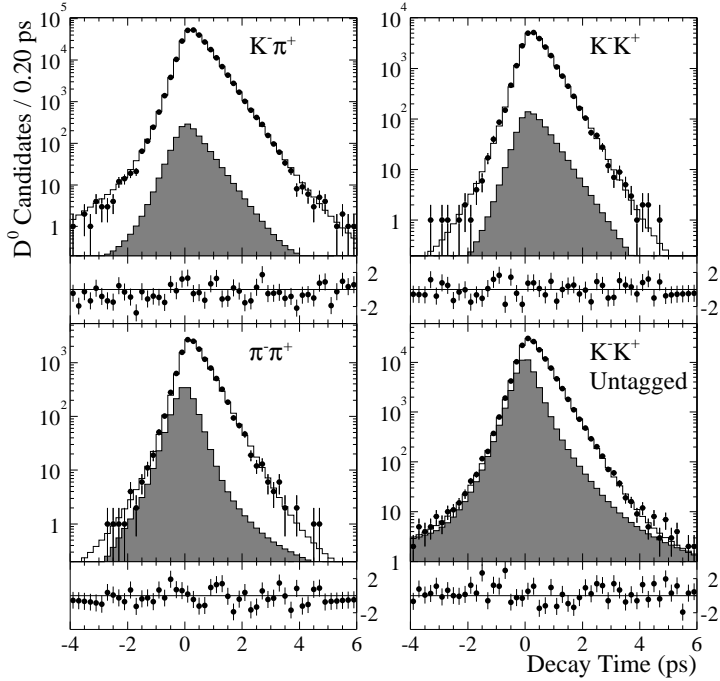,width=3.75in}
\end{center}
\vskip-0.25in
\caption{\it Decay time distributions for 
CF $D^0\ra K^-\pi^+$ (upper left),
$D^0\ra K^+K^-$ (upper right),
$D^0\ra\pi^+\pi^-$ (lower left), and 
$D^0\ra K^+K^-$ selected without using
a $D^{*\,+}$ tag (lower right), from BaBar using 
91~fb$^{-1}$ of data.\cite{cp_babar} The shaded 
histograms show the signal component obtained 
from the fit; residuals from the fit are plotted 
below each distribution.
\label{fig:kk_babar}}
\end{figure}

To date, five experiments\cite{cp_all,cp_babar,cp_focus} have 
measured $y^{}_{CP}$; the most precise value is from BaBar 
using 91~fb$^{-1}$ of data.\cite{cp_babar} To increase statistics, 
BaBar used both $K^+K^-$ and $\pi^+\pi^-$ decays, and, in addition, 
the $D^0\ra K^+K^-$ analysis used both a large inclusive 
$D^0$ sample and a smaller, higher purity sample in which 
the $D^0$ was required to originate from $D^{*+}\ra D^0\pi^+$.
The respective decay time distributions are shown in
Fig.~\ref{fig:kk_babar}.  Doing an unbinned maximum
likelihood fit to each sample, combining results for 
$K^+K^-$ and $\pi^+\pi^-\!$, and taking the ratio of 
lifetimes gives 
$y^{}_{CP}=[ 0.8\,\pm 0.4\,({\rm stat})\,^{+0.5}_{-0.4}\,({\rm syst})]\%$.
This value is consistent with, but smaller than, the 
relatively large value measured by 
FOCUS:\cite{cp_focus} 
$y^{}_{CP}=[ 3.4\,\pm 1.4\,({\rm stat})\,\pm 0.7\,({\rm syst})]\%$. 

BaBar also measures 
$\Delta Y\equiv (\tau^+ -\tau^-)/(\tau^+ + \tau^-)\times 
\tau^{}_{K^-\pi^+}/\langle\tau\rangle$, where $\tau^+\,(\tau^-)$ 
is the lifetime for $D^0\ra K^+K^-$ $(\dbar\ra K^+K^-)$ 
and $\langle\tau\rangle=(\tau^+ + \tau^-)/2$. For $|q/p|\!=\!1$, 
$\Delta Y = x\sin\phi$. The result is 
$\Delta Y\!=\![-0.8\,\pm0.6\,({\rm stat})\,\pm0.2\,({\rm syst})]\%$, 
which indicates that either $x$ is small or $\phi$ is small.

\section{\boldmath $D^0(t)\ra K^+\pi^-$ Doubly-Cabibbo-Suppressed Decays}

For $D^0\ra K^+\pi^-\!$, ${\cal A}^{}_f$ is DCS, $\bar{\cal A}^{}_f$ 
is CF, and thus $|{\cal A}^{}_f|\!\!\ll\!\!|\bar{\cal A}^{}_f|$. In 
addition, there may be a strong phase difference ($\delta$) between 
the amplitudes. Defining 
$R^{}_D\!\equiv\!|{\cal A}^{}_f/\bar{\cal A}^{}_f|^2$ and
$\overline{R}^{}_D\!\equiv\!
|\bar{\cal A}^{}_{\bar{f}}/{\cal A}^{}_{\bar{f}}|^2$,
$\lambda\!=\!|q/p|\,R^{\,-1/2}_D\,e^{i(\phi+\delta)}$ and
$\bar{\lambda}\!=\!|p/q|\,\overline{R}^{\,-1/2}_D\,e^{i(-\phi+\delta)}$.
Inserting these terms into 
Eqs.~(\ref{eqn:master1}) and (\ref{eqn:master2}) gives
\begin{eqnarray}
\hskip-0.30in
R(D^0\ra K^+\pi^-)\!\!\! & \propto\!\!\! &  \nonumber \\
 & & \hskip-1.1in
e^{-\overline{\Gamma}t}
\left[ R^{}_D + \left|\frac{q}{p}\right|\sqrt{R^{}_D}
\left[ y'\cos\phi -x'\sin\phi\right](\overline{\Gamma}t)
+ \left|\frac{q}{p}\right|^2
\frac{(x'^2+y'^2)}{4}(\overline{\Gamma}t)^2\right] 
\label{eqn:kpi_master1} \\
 \nonumber \\
\hskip-0.30in
R(\dbar\ra K^-\pi^+)\!\!\! & \propto\!\!\! &  \nonumber \\
 & & \hskip-1.1in 
e^{-\overline{\Gamma}t}
\left[ \overline{R}^{}_D + \left|\frac{p}{q}\right|\sqrt{\overline{R}^{}_D}
\left[ y'\cos\phi +x'\sin\phi\right](\overline{\Gamma}t)
+ \left|\frac{p}{q}\right|^2
\frac{(x'^2+y'^2)}{4}(\overline{\Gamma}t)^2\right]\,, 
\label{eqn:kpi_master2}
\end{eqnarray}
where $x'\!\equiv\!x\cos\delta +y\sin\delta$ and
$y'\!\equiv\!-x\sin\delta + y\cos\delta$. These 
``rotated'' mixing parameters absorb the unknown 
strong phase difference~$\delta$. \cpv\ enters 
Eqs.~(\ref{eqn:kpi_master1}) and (\ref{eqn:kpi_master2}) 
in three ways: $|q/p|\neq 1$ (\cpv\ in mixing),
$R^{}_D\neq\overline{R}^{}_D$ (\cpv\ in the DCS amplitude),
and $\phi\neq 0$ (\cpv\ via interference between the DCS and
mixed amplitudes). Assuming no \cpv\ gives the simpler
expression
\begin{eqnarray}
R\!\!\! & \propto\!\!\! &  e^{-\overline{\Gamma}t}
\left[ R^{}_D + \sqrt{R^{}_D}\,y'(\overline{\Gamma}t)
+ \frac{(x'^2+y'^2)}{4}(\overline{\Gamma}t)^2\right]\,.
\label{eqn:kpi_master3}
\end{eqnarray}

\begin{figure}[th]
\begin{center}
\vskip0.10in
\hbox{
\epsfig{file=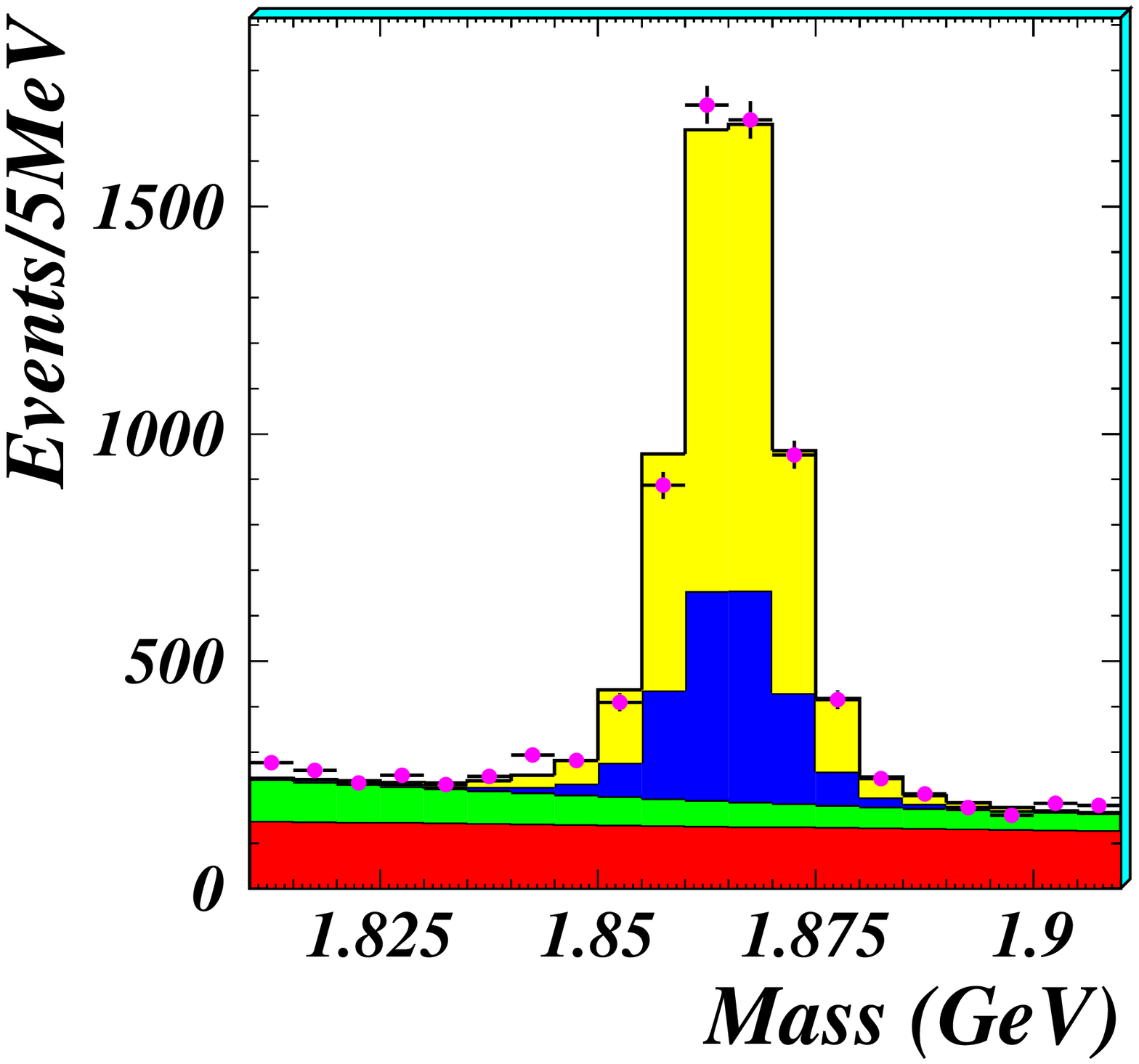,width=2.25in}\hskip0.20in
\epsfig{file=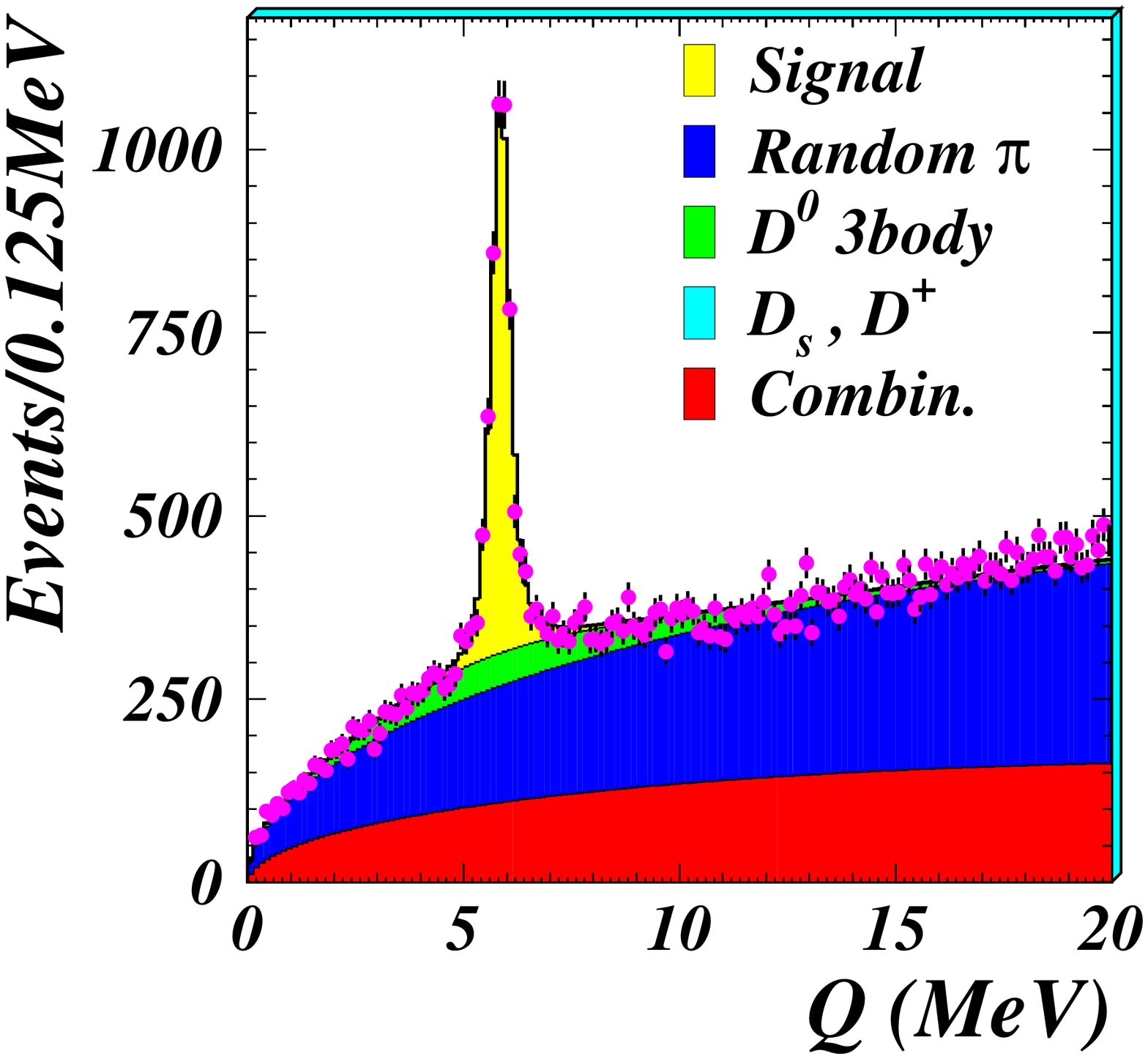,width=2.25in}
}
\end{center}
\vskip-0.35in
\caption{\it 
WS $D^0\ra K^+\pi^-$ decays:
$M^{}_{K\pi}$ spectrum for events satisfying
$Q\!\in(5.3,\,6.5)$~MeV (left), and $Q$
spectrum for events satisfying
$M^{}_{K\pi}\!\in(1.845,\,1.885)$~GeV/$c^2$, from 
Belle using 400~fb$^{-1}$ of data.\cite{kpi_belle} 
\label{fig:kpi_MQ}}
\end{figure}

\begin{figure}[th]
\begin{center}
\vskip0.10in
\hbox{\hskip-0.10in
\epsfig{file=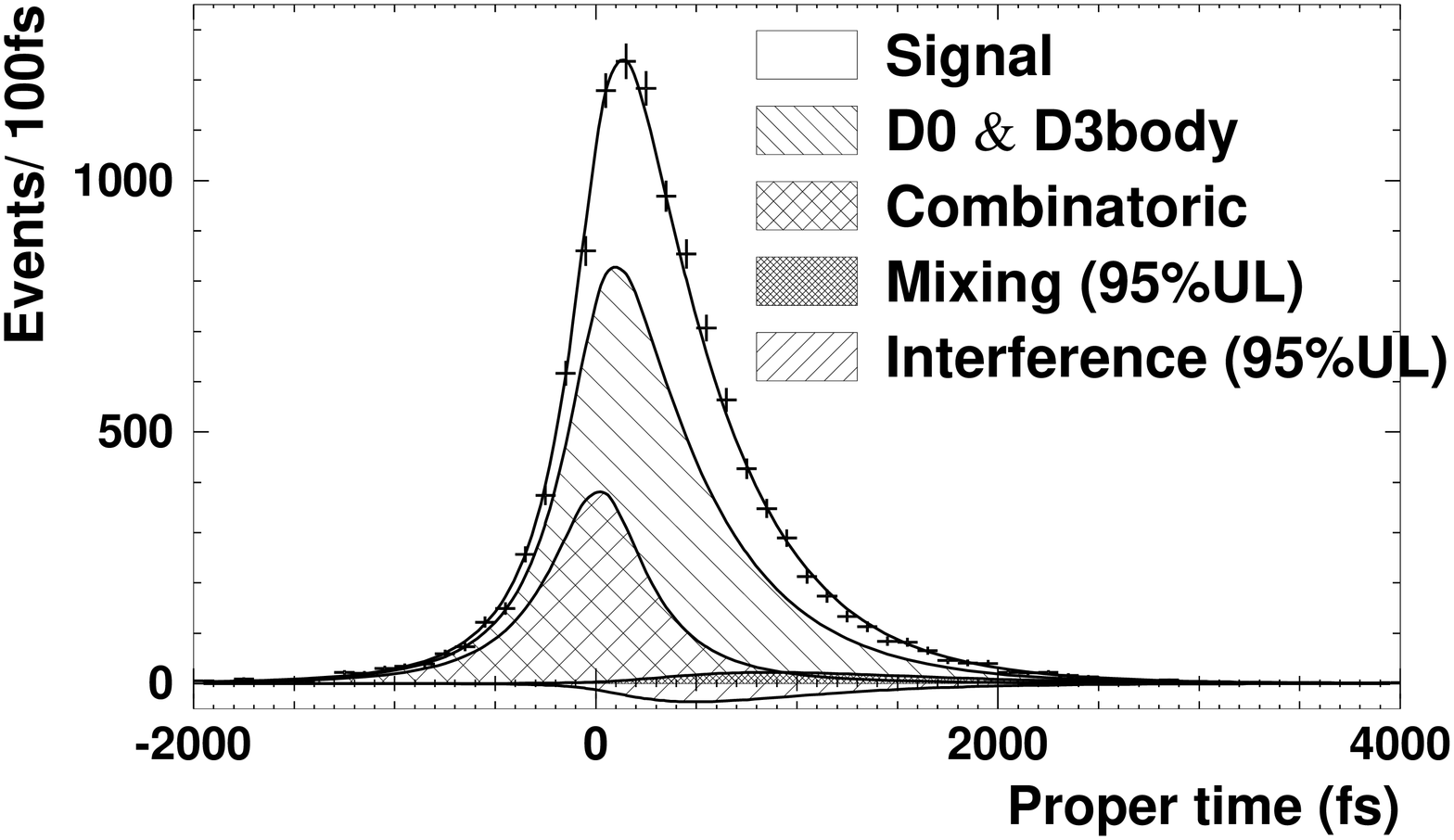,width=2.25in,height=1.90in}
\hskip0.0in
\epsfig{file=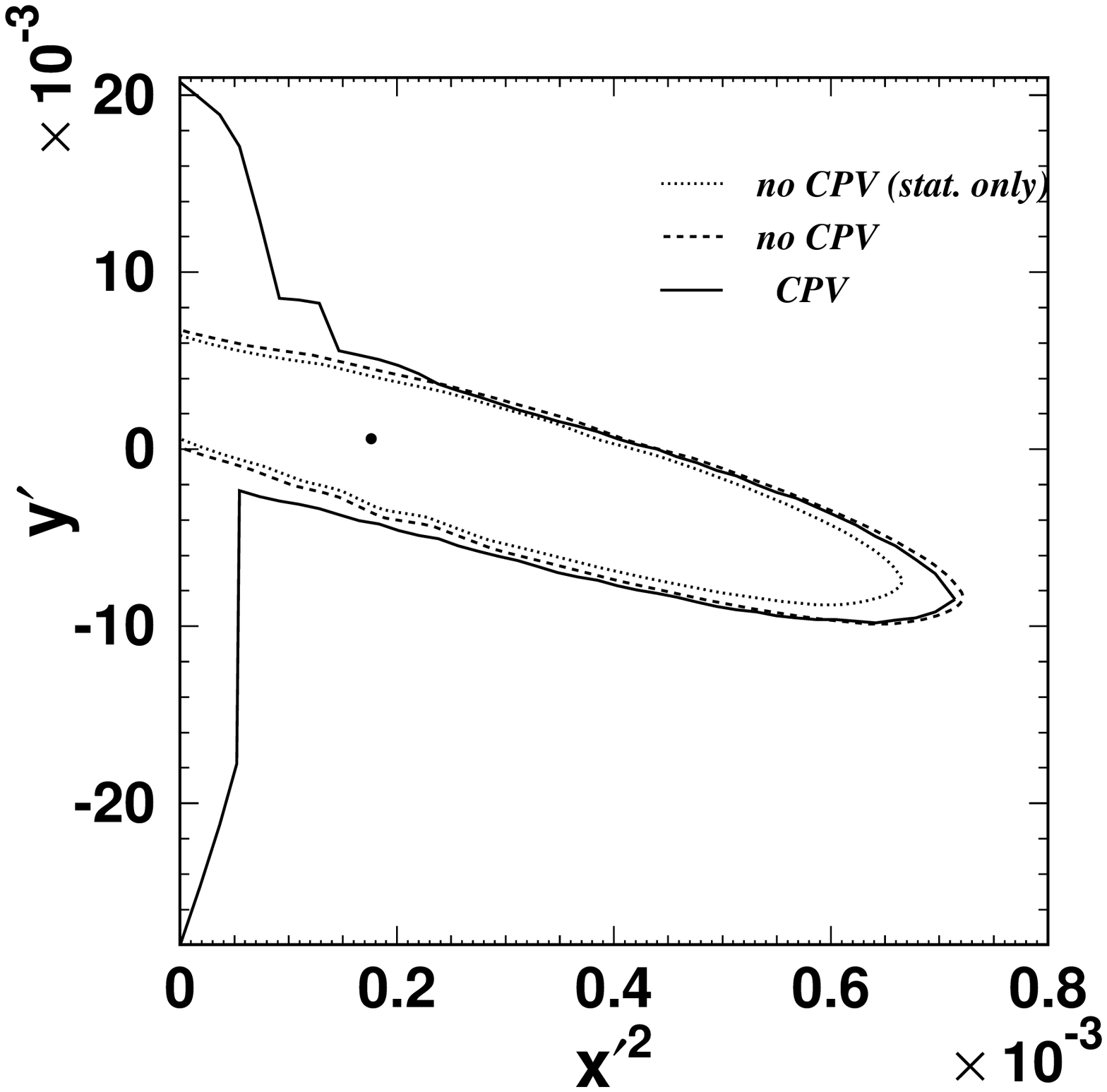,width=2.60in}
}
\end{center}
\vskip-0.35in
\caption{\it Left: WS $D^0\ra K^+\pi^-$ decay-time distribution 
and fit projections. Right: 95\% C.L.\ region for $x'^2$, $y'$.
From Belle using 400~fb$^{-1}$ of data.\cite{kpi_belle}.
\label{fig:kpi_t_contours}}
\end{figure}

To date, six experiments\cite{kpi_e791,kpi_all,kpi_belle} have 
done a time-dependent analysis of $D^0\ra K^+\pi^-$ decays; the 
most stringent constraints on $x'^2$ and $y'$ are 
from Belle using 400~fb$^{-1}$ of data.\cite{kpi_belle}
The reconstructed $M^{}_{K\pi}$ and $Q$ distributions after all
selection criteria are shown in Fig.~\ref{fig:kpi_MQ}; fitting
these distributions yields $1073993\,\pm 1108$ RS signal events
and $4024\,\pm 88$ WS signal events. Those events satisfying
$|M^{}_{K\pi}\!-M^{}_{D^0}|<22$~MeV/$c^2$ and
$|Q-5.8~{\rm MeV}|<1.5$~MeV ($4\sigma$ intervals)
have their decay times fitted for $x'^2,\,y'$, and $R^{}_D$. The 
results are listed in Table~\ref{tab:kpi_results}; projections 
of the fit are shown in Fig.~\ref{fig:kpi_t_contours}(left).

\begin{table}[t]
\centering
\renewcommand{\arraystretch}{1.3}
\caption{\it Limits on mixing parameters obtained from fitting
the decay time distribution of WS $D^0\ra K^+\pi^-$ decays,
from Belle using 400~fb$^{-1}$ of data.\cite{kpi_belle} }
\label{tab:kpi_results}
\vskip 0.1 in
\begin{tabular}{|c|ccc|}
\hline
{\bf Fit Case} & {\bf Parameter} & 
{\bf {\boldmath Fit Result }} &
{\bf {\boldmath 95\% C.L.\ interval }} \\
 & & {\bf {\boldmath $(\times 10^{-3})$}} &
{\bf {\boldmath $(\times 10^{-3})$}} \\
\hline
\hline
No $CPV$ &  
\begin{tabular}{c}
$x'^2$ \\ $y'$ \\ $R^{}_D$ \\ $R^{}_M$ \end{tabular} & 
\begin{tabular}{c}
$0.18\,^{+0.21}_{-0.23}$ \\
$0.6\,^{+4.0}_{-3.9}$ \\
$3.64\pm 0.17$ \\ -- \end{tabular} & 
\begin{tabular}{c}
$<0.72$ \\ 
$(-9.9, 6.8)$ \\
$(3.3, 4.0)$ \\
$(0.63\times\!10^{-5},\,0.40)$ \end{tabular}  \\
\hline
$CPV$ allowed & 
\begin{tabular}{c}
$x'^2$ \\ $y'$  \\ $R^{}_M$ \\ $A^{}_D$ \\ $A^{}_M$ \end{tabular} & 
\begin{tabular}{c}
 -- \\ -- \\ -- \\ 
$23\,\pm 47$ \\ $670\,\pm 1200$ \end{tabular} & 
\begin{tabular}{c}
$< 0.72$  \\ 
$(-28, 21)$ \\
$< 0.40$ \\
$(-76, 107)$ \\
$(-995, 1000)$ \end{tabular} \\
\hline
\hline
No mixing/$CPV$ & $R^{}_D$ & \multicolumn{2}{c|}{ 
$3.77\,\pm 0.08\,({\rm stat})\,\pm 0.05\,({\rm syst})$} \\
\hline
\end{tabular}
\end{table}

A 95\% C.L.\ region in the $x'^2$-$y'$ plane is obtained using 
a frequentist technique based on ``toy'' Monte Carlo (MC) simulation. 
For points $\vec{\alpha}=(x'^2,\,y')$, one generates ensembles
of MC experiments and fits them using the same procedure as that 
used for the data. For each experiment, the difference in likelihood 
$\Delta L\equiv\ln L^{}_{\rm max}\!-\ln L(\vec{\alpha})$ is
calculated, where $L^{}_{\rm max}$ is evaluated for $x'^2\geq 0$.
The locus of points $\vec{\alpha}$ for which 95\% of 
the ensemble has $\Delta L$ less than that of the data 
is taken as the 95\% C.L.\ contour.
This contour is shown in Fig.~\ref{fig:kpi_t_contours}(right);
projections of the contour are listed in the right-most column 
of Table~\ref{tab:kpi_results}.

\cpv\ is accounted for by fitting the $D^0\ra K\pi$ and 
$\dbar\ra K\pi$ samples separately; this yields six values: 
$x'^{2\,\pm},\,y'^\pm$, and~$R^\pm_D$.
Defining $R^\pm_M\!\equiv\!(x'^{\pm\,2}+y'^{\pm\,2})/2$ and
$A^{}_M\!\equiv\!(R^+_M-R^-_M)/(R^+_M+R^-_M)$, one finds
\begin{eqnarray}
x'^\pm & = & \left(\frac{1\pm A^{}_M}{1\mp A^{}_M}\right)^{1/4}
(x'\cos\phi\ \pm\ y'\sin\phi) \label{eqn:result15} 
\label{eqn:cpv1} \\
y'^\pm & = & \left(\frac{1\pm A^{}_M}{1\mp A^{}_M}\right)^{1/4}
(y'\cos\phi\ \mp\ x'\sin\phi)\,,  
\label{eqn:cpv2} 
\end{eqnarray}
where there is an implicit sign ambiguity in $x'^\pm$ due 
to Eqs.~(\ref{eqn:kpi_master1}) and (\ref{eqn:kpi_master2})
being quadratic in $x'$. To allow for \cpv, one obtains separate
$1\!-\!\sqrt{0.05}\!=\!77.6\%$ C.L.\ contours for 
$(x'^{+\,2},\,y'^+)$ and $(x'^{-\,2},\,y'^-)$;
points on the $(x'^{+\,2},\,y'^+)$ contour are 
then combined with points on the $(x'^{-\,2},\,y'^-)$ 
contour and the combination used to solve 
Eqs.~(\ref{eqn:cpv1}) and (\ref{eqn:cpv2})
for $x'^2$ and $y'$. Because the relative sign of
$x'^+$ and $x'^-$ is unknown, there are two
solutions (one for each sign); Belle plots both
in the $(x'^2,\,y')$ plane and takes the outermost 
envelope of points as the 95\% C.L.\ contour allowing 
for~\cpv. This contour has a complicated shape 
[see Fig.~\ref{fig:kpi_t_contours}(right)] due to the two 
solutions. Projections of the contour are listed in the 
right-most column of Table~\ref{tab:kpi_results}. In the case 
of no \cpv, the no-mixing point $x'^2=y'=0$ lies just 
outside the 95\% C.L.\ contour; this point corresponds
to 3.9\%~C.L.\ with systematic uncertainty included.

\section{\boldmath $D^0(t)\ra K^0_S\,\pi^+\pi^-$ Dalitz Plot Analysis}

In this method one considers a self-conjugate final state that 
is {\it not\/} a \cp\ eigenstate, e.g., a three-body decay
that can have either $L\!=\!0$ (\cp-even) or 
$L\!=\!1$ (\cp-odd). If \cpv\ is negligible,  
\cp-eigenstates (denoted $D^{}_-,\,D^{}_+$) 
are mass eigenstates (denoted $D^{}_1,\,D^{}_2$), and
the amplitude for $D^0(t)\ra K^0_S\,\pi^+\pi^-$~is:
\begin{eqnarray}
{\cal A}^{}_{K^0\pi\pi}\!\!\! & =\!\!\! & 
\frac{1}{2p}\Big(
\langle K^0_S\,\pi^+\pi^-|H|D^{}_-(t)\rangle\ +\  
\langle K^0_S\,\pi^+\pi^-|H|D^{}_+(t)\rangle\Big) \nonumber \\
 & \equiv & 
 {\cal A}^{}_-\,e^{-(\Gamma^{}_1/2+im^{}_1)\,t}\ +\ 
 {\cal A}^{}_+\,e^{-(\Gamma^{}_2/2+im^{}_2)\,t} \\
\hskip-0.40in
\Rightarrow\ R^{}_{K^0\pi\pi}\!\!\! & =\!\!\! & 
|{\cal A}^{}_-|^2\,e^{-\overline{\Gamma}(1-y)t}\ +\  
|{\cal A}^{}_+|^2\,e^{-\overline{\Gamma}(1+y)t}\ +\ \nonumber \\
 & & 2e^{-\overline{\Gamma}t}\left[\,
{\rm Re}({\cal A}^{}_+\,{\cal A}^*_-)\cos(\Delta m\,t) +
{\rm Im}({\cal A}^{}_+\,{\cal A}^*_-)\sin(\Delta m\,t)\,\right],
\label{eqn:dalitz_master}
\end{eqnarray}
where ${\cal A}^{}_{+,-}$ is the amplitude for 
$D^{}_{+,-}\!\ra K^0_S\,\pi^+\pi^-$ multiplied by~$1/(2p)$.
Note that $x\!=\!(m^{}_2-m^{}_1)/\overline{\Gamma}$ and
$y\!=\!(\Gamma^{}_2-\Gamma^{}_1)/(2\overline{\Gamma})$. 
For a three-body final state, one can distinguish the ${\cal A}^{}_+$ 
and ${\cal A}^{}_-$ components via a Dalitz plot analysis; 
i.e., a $K^0_S\,f^{}_0(980)$ intermediate state is \cp-even 
and contributes to ${\cal A}^{}_+$, $K^0_S\,\rho^0$ 
is \cp-odd and contributes to ${\cal A}^{}_-$,
$K^*(890)^+\,\pi^-$ is a flavor-eigenstate and contributes to
both ${\cal A}^{}_+$ and ${\cal A}^{}_-$, etc. Thus one models 
${\cal A}^{}_{+,-}$ by separate sums of amplitudes 
$\sum_j a^{}_j\,e^{i\delta^{}_j}A^{}_j$, where $A^{}_j$ is
the Breit-Wigner amplitude\cite{dalitz1_cleo} for resonance
$j$ and is a function of the Dalitz plot position $M^2_{K^0\pi^+}$, 
$M^2_{K^0\pi^-}$. Using the probability density function
of Eq.~(\ref{eqn:dalitz_master}), one does an unbinned maximum 
likelihood fit to $M^2_{K^0\pi^+}$, $M^2_{K^0\pi^-}$, and
the decay time~$t$ to determine $a^{}_j$, $\delta^{}_j$, $x$, 
and $y$. There is systematic uncertainty arising from the 
decay model, i.e., one must decide which intermediate states 
to include in the fit. Unlike Eq.~(\ref{eqn:kpi_master1}), 
Eq.~(\ref{eqn:dalitz_master}) depends linearly on $x$ 
($x\!\ll\!1$) and is therefore sensitive to its sign.

\begin{figure}[th]
\begin{center}
\vskip0.10in
\epsfig{file=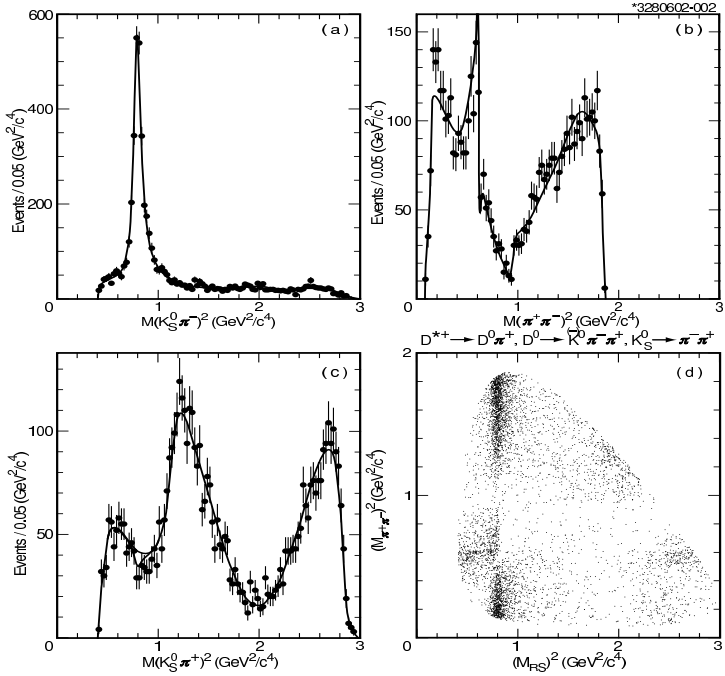,width=3.75in,height=3.5in}
\end{center}
\vskip-0.25in
\caption{\it Dalitz plot (lower right) and projections 
(lower left, upper plots) for $D^0\ra K^0_S\,\pi^+\pi^-$ 
decays, from CLEO using 9.0~fb$^{-1}$ of data.\cite{dalitz3_cleo}
\label{fig:dalitz}}
\end{figure}

This analysis was developed by CLEO, and their 
result\cite{dalitz2_cleo} based on 9.0~fb$^{-1}$ 
has not yet been superseded. To minimize backgrounds,
the $D^0$ candidate is required to originate from $D^{*\,+}\ra D^0\pi^+$.
The final Dalitz plot sample (Fig.~\ref{fig:dalitz})
contains 5299 events with only $(2.1\,\pm 1.5)$\% 
background.\cite{dalitz3_cleo}

The decay model used consists of
$D^0\ra K^*(890)^-\,\pi^+\!$, 
$K^*(1430)^-_{0,2}\,\pi^+\!$, 
$K^*(1680)^-\,\pi^+\!$, 
$K^0_S\,\rho$, 
$K^0_S\,\omega$, 
$K^0_S\,f^{}_0(980)$, 
$K^0_S\,f^{}_2(1270)$, 
$K^0_S\,f^{}_0(1370)$, 
WS $D^0\ra K^*(890)^+\,\pi^-\!$, 
and a nonresonant component.
The fit results are listed in Table~\ref{tab:dalitz}; the
95\%~C.L.\ intervals correspond to the values at which
$-2\ln{\cal L}$ rises by 3.84 units, where ${\cal L}$ is 
the likelihood function. \cpv\ is included in the fit by 
introducing parameters $\varepsilon\equiv (p-q)/(p+q)$ 
(in analogy with $K^0$ decays) and $\phi$, the weak phase 
difference between $\bar{\cal A}^{}_{K^0\pi\pi}$ 
and~${\cal A}^{}_{K^0\pi\pi}$. The results listed are 
consistent with no mixing or~\cpv.

\begin{table}[t]
\centering
\renewcommand{\arraystretch}{1.2}
\caption{\it Limits on mixing and \cpv\ parameters from a 
$t$-dependent fit to the $D^0\ra K^0_S\,\pi^+\pi^-$ Dalitz 
plot, from CLEO using 9.0~fb$^{-1}$.\cite{dalitz2_cleo} 
The errors are statistical, experimental systematic,
and modeling systematic, respectively. }
\label{tab:dalitz}
\vskip 0.1 in
\begin{tabular}{|c|ccc|}
\hline
{\bf Fit} & {\bf Param.} & {\bf Fit Result\,(\%)} &
{\bf {\boldmath 95\% C.L.\,Inter.\,(\%)}} \\
\hline
\hline
No $CPV$ 
&  
\begin{tabular}{c}
$x$ \\ $y$ \\
\end{tabular} 
& 
\begin{tabular}{c}
$1.8\,^{+3.4}_{-3.2}\,\pm 0.4\,\pm 0.4$  \\
$-1.4\,^{+2.5}_{-2.4}\,\pm 0.8\,\pm 0.4$ \\
\end{tabular} 
& 
\begin{tabular}{c}
$(-4.7, 8.6)$ \\ 
$(-6.3, 3.7)$ \\
\end{tabular}
\\ 
\hline\hskip-0.20in
\begin{tabular}{c}
$CPV$ \\ ~~Allowed 
\end{tabular}
&  
\begin{tabular}{c}
$x$ \\ $y$ \\ $\epsilon$ \\ $\phi$  \\ 
\end{tabular} 
& 
\begin{tabular}{c}
$2.3\,^{+3.5}_{-3.4}\,\pm 0.4\,\pm 0.4$  \\
$-1.5\,^{+2.5}_{-2.4}\,\pm 0.8\,\pm 0.4$ \\
$1.1\,\pm 0.7\,\pm 0.4\,\pm 0.2$ \\
$(5.7\,\pm 2.8\,\pm 0.4\,\pm 1.2)^\circ$ \\
\end{tabular} 
& 
\begin{tabular}{c}
$(-4.5, 9.3)$ \\ 
$(-6.4, 3.6)$ \\ 
$(-0.4, 2.4)$ \\ 
$(-0.3^\circ, 11.7^\circ)$ \\
\end{tabular}
\\ 
\hline
\end{tabular}
\end{table}

\section{\boldmath $D^0(t)\ra K^+\pi^-\pi^0$ and $K^+\pi^-\pi^+\pi^-$
  Multibody Decays}

\begin{figure}[thb]
\begin{center}
\epsfig{file=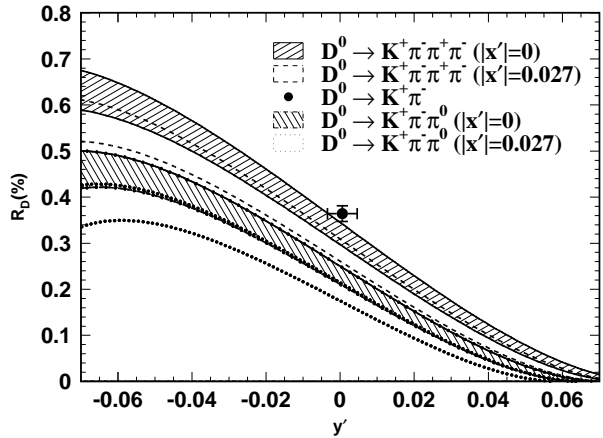,width=3.1in}
\end{center}
\vskip-0.25in
\caption{\it WS $D^0\ra K^+\pi^- n(\pi)$ decays:
95\% C.L.\ range for $R^{}_D$ as a function of $y'$ for 
$|x'|=0$ and $|x'|=0.027$, from Belle using 281~fb$^{-1}$ 
of data.\cite{multi_belle} The point with $1\sigma$ error
bars is Belle's result from $D^0\ra K^+\pi^-$ decays
(see Table~\ref{tab:kpi_results}).
\label{fig:multi}}
\end{figure}

Mixing has also been searched for in WS multibody 
final states\cite{kpi_e791,multi_all,multi_belle}  
$K^+\pi^-\pi^0$ and $K^+\pi^-\pi^+\pi^-$; the most recent 
measurement is from Belle using 281 fb$^{-1}$ of 
data.\cite{multi_belle} The final signal yields 
are $1978\pm 104$ $D^0\ra K^+\pi^-\pi^0$ decays and 
$1721\pm 75$ $D^0\ra K^+\pi^-\pi^+\pi^-$ decays.
For this analysis no decay time information is used, 
i.e., Belle measures the time-integrated ratio of 
WS to RS decays: 
\begin{eqnarray}
R^{}_{\rm WS}\!\!\! & =\!\!\! & \frac{\int R[D^0\ra K^+\pi^- (n\pi)]\,dt}
{\int R[D^0\ra K^-\pi^+ (n\pi)]\,dt}
\ \approx\ R^{}_D + \sqrt{R^{}_D}\,y' + \frac{x'^2+y'^2}{2}\,,
\label{eqn:multi_master}
\end{eqnarray}
where $R^{}_D$ is the ratio of the DCS rate to the CF rate
as previously defined for $D^0\ra K^+\pi^-$ decays. The results 
are $R^{}_{\rm WS}=
[ 0.229\,\pm 0.015\,({\rm stat})\,^{+0.013}_{-0.009}\,({\rm syst})]$\%
for $K^+\pi^-\pi^0$ and 
$[ 0.320\,\pm 0.018\,({\rm stat})\,^{+0.018}_{-0.013}\,({\rm syst})]$\%
for $K^+\pi^-\pi^+\pi^-$. Inserting these values into 
Eq.~(\ref{eqn:multi_master}) allows one to determine 
$R^{}_D$ as a function of $x'$ or~$y'$.
Assuming $x'\!=\!0$ and $|x'|\!=\!0.027$ 
gives the curves shown in Fig.~\ref{fig:multi}; the latter 
$|x'|$ value corresponds to Belle's 95\% C.L.\ upper limit 
from $D^0\ra K^+\pi^-$ decays (see Table~\ref{tab:kpi_results}).
However, the value of $x'$ from $D^0\ra K^+\pi^-$ may differ 
from that from $D^0\ra K^+\pi^- n(\pi)$ due to the strong phase
differences ($\delta$) being different.

\section{Summary}

\begin{figure}[thb]
\begin{center}
\epsfig{file=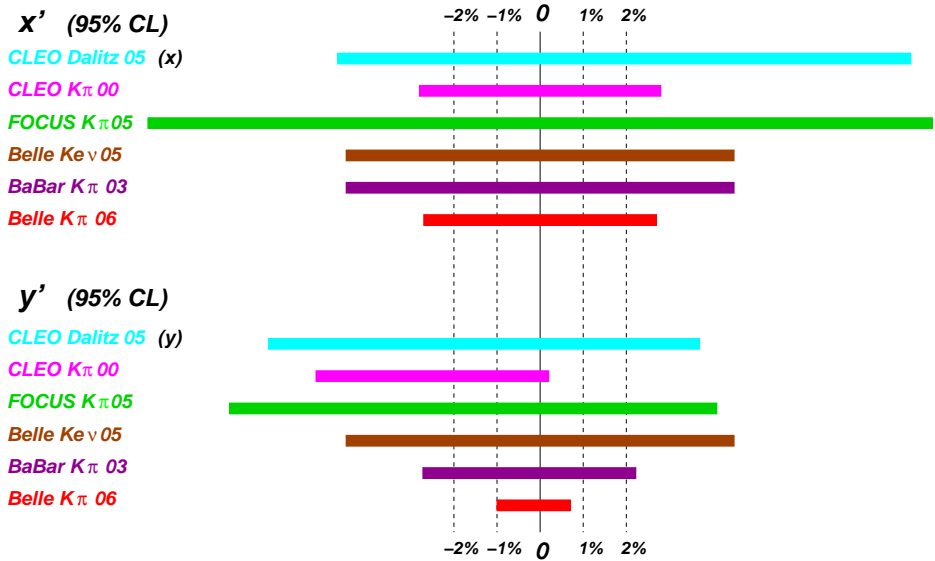,width=4.85in}
\end{center}
\vskip-0.20in
\caption{\it 95\% C.L.\ allowed ranges for $x'$ (top)
and $y'$ (bottom) from various experiments assuming 
no~\cpv. The CLEO Dalitz results are for $x$ and~$y$.
\label{fig:summary_xy}}
\end{figure}

The 95\% C.L.\ allowed ranges for $x'$ and $y'$ are plotted 
in Fig.~\ref{fig:summary_xy}; for simplicity we assume 
negligible~\cpv. The most stringent constraints are 
$|x'|\!<\!2.7$\% and $y'\!\in(-1.0\%,\,0.7\%)$.
These ranges are projections of the two-dimensional 
95\%~C.L.\ region for $x'^2,\,y'$ from Belle
[Fig.~\ref{fig:kpi_t_contours}(right)].

\begin{figure}[bht]
\begin{center}
\epsfig{file=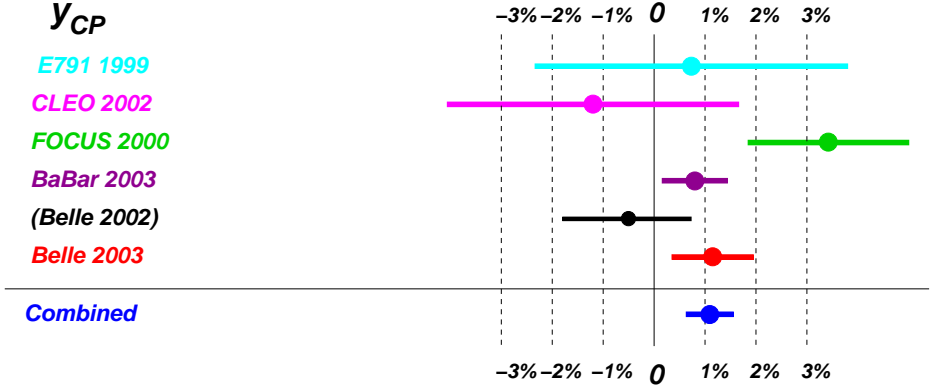,width=4.85in}
\end{center}
\vskip-0.25in
 \caption{\it $y^{}_{CP}$ central values and $1\sigma$ 
errors measured by various experiments, and the combined
result assuming the individual errors uncorrelated. The 
Belle 2002 data sample (23~fb$^{-1}$) has some overlap with 
the Belle 2003 data sample (158~fb$^{-1}$), and thus this
result is not included in the average.
\label{fig:summary_ycp}}
\end{figure}

The results for $y^{}_{CP}$ are plotted in 
Fig.~\ref{fig:summary_ycp}. Here the central values and 
$1\sigma$ errors are shown; combining the results assuming 
the errors uncorrelated gives $y^{}_{CP}\!=\!(1.09\,\pm 0.46)$\%. 
This value differs from zero by $2.4 \sigma$ and indicates 
a nonzero decay width difference $\Delta\Gamma$. Assuming 
negligible \cpv, one can combine this value with Belle's 
central value for $y'$, $(0.06\,^{+0.40}_{-0.39})\%$. The result 
is $y'/y=\cos\delta-(x/y)\sin\delta =
0.05\,^{+0.39}_{-0.37}$, where the error is obtained from 
an MC calculation as the fractional errors on $y$ and $y'$ 
are large. This small central value (albeit with a large error) 
implies $\tan\delta\approx y/x$; i.e., if $x\!\ll\!y$, 
then $\delta$ is near 90$^\circ$. Such a strong phase difference 
would be much larger than expected.

\end{document}